\documentstyle[aps,prl,psfig,multicol]{revtex}
\newcommand{\be}{\begin{equation}}
\newcommand{\ee}{\end{equation}}
\newcommand{\bes}{\begin{eqnarray}}
\newcommand{\ees}{\end{eqnarray}}
\begin{document}
\bibliographystyle{prsty}
\title{Local estimates for entropy densities in coupled map lattices}
\author{Eckehard Olbrich, Rainer Hegger and Holger Kantz}
\address{Max-Planck-Institut f\"ur Physik komplexer Systeme\\
N\"othnitzer  Str. 38, D 01187 Dresden, Germany \\ 
e-mail: olbrich@mpipks-dresden.mpg.de \\
tel.: +49-351-871-1206 \\
fax: +351-871-1199} 

\maketitle

\begin{abstract}
We present a method to derive an upper bound for the
entropy density of coupled map lattices with local interactions from
local observations. To do this, we use an embedding technique being
a combination of time delay and spatial embedding. This embedding allows
us to identify the local character of the equations of motion.
Based on this method we present an approximate estimate of the
entropy density by the correlation integral.
\end{abstract}


\begin{multicols}{2}
Many natural phenomena involving a huge number of degrees of freedom
emerge from spatially extended systems (SESs), with
translationally invariant equations of motion and spatial coupling.
Thus an understanding of these kinds of systems is vital for the
understanding of nature. Only for very few of them there exist models
from first principles which allow us to explore them by analytical or
numerical means. Many systems can not be satisfactorily modeled, so
that one has to rely on the study of experimental observations. This
usually leads to time series analysis, which has gained new stimulus
in recent years through the concept of phase space reconstruction.
This approach is quite successful for low
dimensional chaotic systems~\cite{Abarbanel96,Kantz97c}. There one can obtain
characteristic quantities like Lyapunov exponents, fractal dimensions
and entropies directly from the measurements. In some cases it
is even possible to construct reasonable model equations from the data.

Natural SESs are typically continuous in time and space and thus
described by partial differential equations (PDEs). Nevertheless, a
PDE can be approximated by a coupled map lattice (CML), i.e. a model
discrete in time and space. CMLs were used as paradigmatic
models to study general properties of spatio-temporal chaotic
systems. A fundamental question concerning SESs is whether it is
possible to extract characteristic quantities from local
measurements. This problem is still unsolved
in the realm of time series analysis.
There were a considerable number of attempts,
especially for the dimension density
\cite{Mayer-Kress87b,Mayer-Kress89a,Grassberger89,Torcini91,Korzinov92,Bauer93,Tsimring93}.
Nevertheless, the outcome was far from being satisfactory. Common to all
these works is to use either time delay or purely spatial embedding for
reconstructing the dynamical states or the invariant
measure. The general problem is that local measurements can only
reconstruct finite dimensional subspaces, though the whole phase space
is infinite dimensional, in principle. Therefore, these subsystems are
never fully deterministic, since they are coupled to the unobserved
part of the phase space and are thus open systems.

However, in systems with local, e.g. nearest neighbour, interactions,
the equations of motion include only a few local
variables. Regarding only these variables the ``local`` future is
governed by deterministic laws, while the whole dynamics in the
measured subspace remains ''stochastically'' driven. The main idea we want
to present is that the local deterministic structure can be exploited
to estimate a dynamical entropy which will turn out to be a good
approximation for the entropy density of the whole (mainly
unobserved) system. 

The general CML we want to study here is given by
\begin{eqnarray}
x_i(n+1)&=&(1-\sigma_r-\sigma_l)f(x_i(n))\nonumber\\
&&+\sigma_lg(x_{i-1}(n))+\sigma_rg(x_{i+1}(n))\;,
\label{cml-def}
\end{eqnarray}
where $i$ represents the position in space and $n$ the time. $\sigma_r$
and $\sigma_l$ are spatial coupling constants. Concerning the coupling,
typically two kinds of systems are 
studied: (1) $\sigma_r=\sigma_l=\sigma$ which we will refer to as the
(symmetric) diffusive coupling and (2)
$\sigma_{r(l)}=0\;,\sigma_{l(r)}=\sigma$ the unidirectional
coupling. The map $f$ is a chaotic map which describes the local
dynamics. We will use the tent map $f(x)=1-2 |x-1/2|$ throughout this
paper. $g$ is an in principle arbitrary
function. We use either $g=f$ which gives rise to a nonlinear coupling
or $g(x)=x$ which results in a linear coupling.

It is commonly observed that the dynamics of these CMLs is 
extensively chaotic in suitable parameter regions. Extensive means
that the attractor dimension and the KS-entropy are proportional
to the system size, here the number $N$ of lattice sites. Thus it is
possible to define dimension and entropy densities as intensive
quantities. Usually these densities are estimated via the Pesin identity
or the Kaplan-York formula, respectively, by calculating the
Lyapunov spectrum using the model equations. In some cases this is
even possible if the equations of motion are unknown~\cite{Buenner99}.

In previous attempts to calculate entropies and dimensions from
observed time series either pure spatial embedding with state
vectors of the form $(x_i(n),\ldots,x_{i+l-1}(n))$ or temporal
embedding with the vectors $(x_i(n),\ldots,x_i(n+m-1))$ were used. Since the
dynamics for all these embeddings is non-autonomous with respect to
equation (\ref{cml-def}), the corresponding projections of the
invariant measure of the CML contain stochastic components so that the
entropy diverges \cite{Gaspard93c}. In this letter we introduce a
novel embedding procedure (``pyramid embedding'')
which does not increase the dimension of the
reconstructed measure when the dimension of the involved subspace is
increased. This will enable us to compute the finite entropy related
to the deterministic part of the dynamics in the subspace.

Consider a partitioning of a subspace $\Gamma_{\vec{s}}$ spanned by the
components of a state vector $\vec{s}$ with a rectangular grid
of mesh size $\epsilon$. Via the invariant measure we can assign a probability
$p_i$ to every cell and define the entropy 
\be
H(\vec{s},\epsilon)=-\sum_i p_i \ln p_i \; .
\label{def-shannon}
\ee
For sufficient small $\epsilon$ its $\epsilon$-dependence is given by
\cite{Gaspard93c}
\be
H(\vec{s},\epsilon) \propto -D \ln \epsilon\; ,
\ee
where $D$ is the information dimension of the projection of the
invariant measure into the subspace spanned by $\vec{s}$.
Let us choose a second subspace $\Gamma_{\vec{t}}$ spanned by the components of
another state vector $\vec{t}$ satisfying
\be
\vec{t}=\vec{F}(\vec{s}\,) \;,
\label{constraint}
\ee
which implies that the constraint $\vec{F}$ is determined by equation
(\ref{cml-def}). $H(\vec{t},\vec{s},\epsilon)$ denotes the
entropy of the joint probability $p_{ij}$ for the system being in
cell $i$ of $\Gamma_{\vec{s}}$ and in cell $j$ of $\Gamma_{\vec{t}}$.
The projection of the invariant measure in the enlarged state
space has the identical information dimension $D$ due to the
constraints (\ref{constraint}). 
Therefore the conditional entropy
\be
h(\vec{t}\,|\vec{s}\,,\epsilon)=
H(\vec{t},\vec{s}\,,\epsilon)-H(\vec{s}\, ,\epsilon)
\ee
will become
independent of $\epsilon$  for sufficient small $\epsilon$.

In the following we will use a abbreviated 
symbolic representation for the states appearing as arguments in the
entropies. The pure spatial state of $l$ neighbouring sites will be
denoted by
\[
\left( x_1(n),x_2(n),\ldots,x_{l-1}(n),x_{l}(n)\right) \rightarrow
\underbrace{ 
\begin{picture}(12,2) \thinlines
\put(0,0){\line(1,0){4}} \put(0,2){\line(1,0){4}} 
\put(0,0){\line(0,1){2}}  \put(2,0){\line(0,1){2}}
\put(4,0){\line(0,1){2}}
\multiput(5,1)(1,0){3}{\circle*{0.5}} 
\put(8,0){\line(1,0){4}} \put(8,2){\line(1,0){4}} 
\put(8,0){\line(0,1){2}}  \put(10,0){\line(0,1){2}} \put(12,0){\line(0,1){2}}
\end{picture} }_{l} \; .
\] 
This notation will allow us to write states
which combine spatial and temporal embedding in a compact way.
Note that we can omit the time and space indices because of the
stationarity and translation invariance.
 
The simplest way to choose state vectors
$\vec{s}$ and $\vec{t}$ fulfilling the constraint (\ref{constraint})
is $\vec{s}=(x_i(n),x_{i+1}(n))$ and $\vec{t}=(x_i(n+1))$ 
for unidirectional coupling and $\vec{s}=(x_{i-1}(n),x_i(n),x_{i+1}(n))$ and
$\vec{t}=(x_i(n+1))$ for diffusive coupling, respectively.
Using the symbolic writing we get for the unidirectional case
\begin{equation}
 H(\vec{s}\, ,\epsilon)=:H \left( \begin{picture}(4,2) \thinlines
\put(0,0){\line(1,0){4}} \put(0,2){\line(1,0){4}} 
\put(0,0){\line(0,1){2}}  \put(2,0){\line(0,1){2}} \put(4,0){\line(0,1){2}} 
\end{picture} \right) \; \;
\mbox{and} \; \;
H(\vec{t},\vec{s}\, , \epsilon)=:H \left( \begin{picture}(4,4) \thinlines
\put(0,-1){\line(1,0){4}} \put(0,1){\line(1,0){4}} \put(0,3){\line(1,0){2}}
\put(0,-1){\line(0,1){4}}  \put(2,-1){\line(0,1){4}} \put(4,-1){\line(0,1){2}} 
\end{picture} \right) \;. \label{eq.above}
\end{equation}
In the symbolic notation of conditional entropies
$h(\vec{t}|\vec{s})$ we will hatch $\vec t$, e.g. (\ref{eq.above}) yields
\[
h(\vec{t}\,|\vec{s}\, , \epsilon)=H(\vec{t},\vec{s}\, , \epsilon)
-H(\vec{s}\, , \epsilon) =: h \left( \begin{picture}(4,4) \thinlines
\put(0,-1){\line(1,0){4}} \put(0,1){\line(1,0){4}}
\put(0,1.5){\line(1,0){2}} \put(0,2){\line(1,0){2}} \put(0,2.5){\line(1,0){2}}
\put(0,3){\line(1,0){2}}
\put(0,-1){\line(0,1){4}}  \put(2,-1){\line(0,1){4}} \put(4,-1){\line(0,1){2}} 
\put(0.5,1){\line(0,1){2}} \put(1,1){\line(0,1){2}} \put(1.5,1){\line(0,1){2}}
\end{picture} \right) \; .
\]

The KS-entropy of a dynamical system is defined as the conditional
entropy of the state of the system knowing the full past. 
Because the
KS-entropy is proportional to the system size $N$ in our case, we can
introduce the entropy density $\eta$ which is the KS-entropy divided
by $N$.
The definition of the entropy density can be written as
\be
\eta=\lim_{N\to\infty} \lim_{m \to \infty} \lim_{\epsilon \to 0}
\frac{1}{N} \left( h \left( ^{\normalsize m+1 \Big\{} \underbrace{ \begin{picture}(12,10) \thinlines
\put(0,0){\line(1,0){4}} \put(0,2){\line(1,0){4}} 
\put(0,0){\line(0,1){2}}  \put(2,0){\line(0,1){2}}
\put(4,0){\line(0,1){2}}
\multiput(5,1)(1,0){3}{\circle*{0.5}}
\put(8,0){\line(1,0){4}} \put(8,2){\line(1,0){4}} 
\put(8,0){\line(0,1){2}}  \put(10,0){\line(0,1){2}}
\put(12,0){\line(0,1){2}}
\multiput(1,3)(1,0){3}{\circle*{0.5}}
\multiput(5,3)(1,0){3}{\circle*{0.5}}
\multiput(9,3)(1,0){3}{\circle*{0.5}}
\put(0,4){\line(1,0){4}} \put(0,6){\line(1,0){4}} 
\put(0,8){\line(1,0){4}}
\put(0,4){\line(0,1){4}}  \put(2,4){\line(0,1){4}}
\put(4,4){\line(0,1){4}}
\put(0,6.5){\line(1,0){4}}
\put(0,7){\line(1,0){4}}
\put(0,7.5){\line(1,0){4}}
\multiput(5,7)(1,0){3}{\circle*{0.5}}
\multiput(5,5)(1,0){3}{\circle*{0.5}}
\put(8,4){\line(1,0){4}} \put(8,6){\line(1,0){4}}
\put(8,8){\line(1,0){4}}
\put(8,6.5){\line(1,0){4}}
\put(8,7){\line(1,0){4}}
\put(8,7.5){\line(1,0){4}}
\put(8,4){\line(0,1){4}}  \put(10,4){\line(0,1){4}}
\put(12,4){\line(0,1){4}}
\put(0.5,6){\line(0,1){2}}
\put(1,6){\line(0,1){2}}
\put(1.5,6){\line(0,1){2}}
\put(2.5,6){\line(0,1){2}}
\put(3,6){\line(0,1){2}}
\put(3.5,6){\line(0,1){2}}
\put(8.5,6){\line(0,1){2}}
\put(9,6){\line(0,1){2}}
\put(9.5,6){\line(0,1){2}}
\put(10.5,6){\line(0,1){2}}
\put(11,6){\line(0,1){2}}
\put(11.5,6){\line(0,1){2}}
\end{picture} 
 }_{N} \right) \right) \; ,
\label{eta2}
\ee
where the entropies are
calculated for a finite CML with N lattice sites. 
We rewrite the r.h.s. of (\ref{eta2}) in such a way that only
a single site remains as the conditioned part in the conditional
entropies \cite{footnote}:
\be
h \left( \underbrace{ \begin{picture}(12,10) \thinlines
\put(0,0){\line(1,0){4}} \put(0,2){\line(1,0){4}} 
\put(0,0){\line(0,1){2}}  \put(2,0){\line(0,1){2}}
\put(4,0){\line(0,1){2}}
\multiput(5,1)(1,0){3}{\circle*{0.5}}
\put(8,0){\line(1,0){4}} \put(8,2){\line(1,0){4}} 
\put(8,0){\line(0,1){2}}  \put(10,0){\line(0,1){2}}
\put(12,0){\line(0,1){2}}
\multiput(1,3)(1,0){3}{\circle*{0.5}}
\multiput(5,3)(1,0){3}{\circle*{0.5}}
\multiput(9,3)(1,0){3}{\circle*{0.5}}
\put(0,4){\line(1,0){4}} \put(0,6){\line(1,0){4}} 
\put(0,8){\line(1,0){4}}
\put(0,4){\line(0,1){4}}  \put(2,4){\line(0,1){4}}
\put(4,4){\line(0,1){4}}
\put(0,6.5){\line(1,0){4}}
\put(0,7){\line(1,0){4}}
\put(0,7.5){\line(1,0){4}}
\multiput(5,7)(1,0){3}{\circle*{0.5}}
\multiput(5,5)(1,0){3}{\circle*{0.5}}
\put(8,4){\line(1,0){4}} \put(8,6){\line(1,0){4}}
\put(8,8){\line(1,0){4}}
\put(8,6.5){\line(1,0){4}}
\put(8,7){\line(1,0){4}}
\put(8,7.5){\line(1,0){4}}
\put(8,4){\line(0,1){4}}  \put(10,4){\line(0,1){4}}
\put(12,4){\line(0,1){4}}
\put(0.5,6){\line(0,1){2}}
\put(1,6){\line(0,1){2}}
\put(1.5,6){\line(0,1){2}}
\put(2.5,6){\line(0,1){2}}
\put(3,6){\line(0,1){2}}
\put(3.5,6){\line(0,1){2}}
\put(8.5,6){\line(0,1){2}}
\put(9,6){\line(0,1){2}}
\put(9.5,6){\line(0,1){2}}
\put(10.5,6){\line(0,1){2}}
\put(11,6){\line(0,1){2}}
\put(11.5,6){\line(0,1){2}}
\end{picture} 
 }_{N} \right) = h \left( \underbrace{ \begin{picture}(12,10) \thinlines
\put(0,0){\line(1,0){4}} \put(0,2){\line(1,0){4}} 
\put(0,0){\line(0,1){2}}  \put(2,0){\line(0,1){2}}
\put(4,0){\line(0,1){2}}
\multiput(5,1)(1,0){3}{\circle*{0.5}}
\put(8,0){\line(1,0){4}} \put(8,2){\line(1,0){4}} 
\put(8,0){\line(0,1){2}}  \put(10,0){\line(0,1){2}}
\put(12,0){\line(0,1){2}}
\multiput(1,3)(1,0){3}{\circle*{0.5}}
\multiput(5,3)(1,0){3}{\circle*{0.5}}
\multiput(9,3)(1,0){3}{\circle*{0.5}}
\put(0,4){\line(1,0){4}} \put(0,6){\line(1,0){4}} 
\put(0,8){\line(1,0){2}}
\put(0,4){\line(0,1){4}}  \put(2,4){\line(0,1){4}}
\put(4,4){\line(0,1){2}}
\put(0,6.5){\line(1,0){2}}
\put(0,7){\line(1,0){2}}
\put(0,7.5){\line(1,0){2}}
\multiput(5,5)(1,0){3}{\circle*{0.5}}
\put(8,4){\line(1,0){4}} \put(8,6){\line(1,0){4}}
\put(8,4){\line(0,1){2}}  \put(10,4){\line(0,1){2}}
\put(12,4){\line(0,1){2}}
\put(0.5,6){\line(0,1){2}}
\put(1,6){\line(0,1){2}}
\put(1.5,6){\line(0,1){2}}
\end{picture} 
}_{N} \right) 
+ \ldots + h \left( \underbrace{  \begin{picture}(12,10) \thinlines
\put(0,0){\line(1,0){4}} \put(0,2){\line(1,0){4}} 
\put(0,0){\line(0,1){2}}  \put(2,0){\line(0,1){2}}
\put(4,0){\line(0,1){2}}
\multiput(5,1)(1,0){3}{\circle*{0.5}}
\put(8,0){\line(1,0){4}} \put(8,2){\line(1,0){4}} 
\put(8,0){\line(0,1){2}}  \put(10,0){\line(0,1){2}}
\put(12,0){\line(0,1){2}}
\multiput(1,3)(1,0){3}{\circle*{0.5}}
\multiput(5,3)(1,0){3}{\circle*{0.5}}
\multiput(9,3)(1,0){3}{\circle*{0.5}}
\put(0,4){\line(1,0){4}} \put(0,6){\line(1,0){4}} 
\put(0,8){\line(1,0){4}}
\put(0,4){\line(0,1){4}}  \put(2,4){\line(0,1){4}}
\put(4,4){\line(0,1){4}}
\multiput(5,7)(1,0){3}{\circle*{0.5}}
\multiput(5,5)(1,0){3}{\circle*{0.5}}
\put(8,4){\line(1,0){4}} \put(8,6){\line(1,0){4}}
\put(8,8){\line(1,0){4}}
\put(10,6.5){\line(1,0){2}}
\put(10,7){\line(1,0){2}}
\put(10,7.5){\line(1,0){2}}
\put(8,4){\line(0,1){4}}  \put(10,4){\line(0,1){4}}
\put(12,4){\line(0,1){4}}
\put(10.5,6){\line(0,1){2}}
\put(11,6){\line(0,1){2}}
\put(11.5,6){\line(0,1){2}}
\end{picture} 
}_{N} \right)
\label{identity}
\ee
The translational invariance and periodic boundary
conditions allow cyclic permutations of columns. 
After shifting $\vec t$ to the leftmost column  in each single term in
(\ref{identity}), we split the blocks $\vec s$ into
$\vec s_1$ and $\vec s_2$ like
\[
\psfig{file=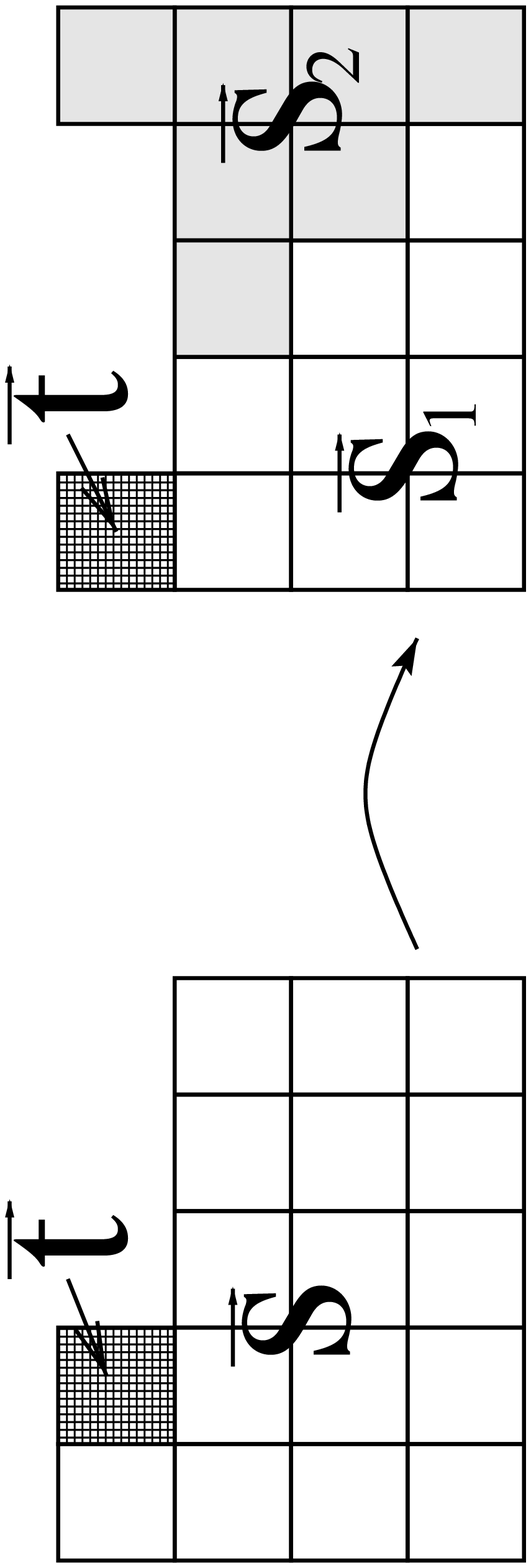,width=4cm,angle=270}
\]
For any $\vec{s_2}$, $h(\vec{t}\,|\vec{s_1}\, \vec{s_2})$ fulfills the
inequality
\be
h(\vec{t}\,|\vec{s_1}\,) \ge h(\vec{t}\,|\vec{s_1} \vec{s_2}) 
\label{inequality}
\ee
It says
that the uncertainty about the state $\vec{t}$ is the larger the less
I know about the rest of the system. Formally this can be shown using Jensen's
inequality (see e.g. \cite{Cover91}).
If we apply (\ref{inequality}) to (\ref{identity}) we get
\be
h \left( ^{{\scriptstyle{m+1}} \Big\{} \;  \underbrace{ \begin{picture}(12,10) \thinlines
\put(0,0){\line(1,0){4}} \put(0,2){\line(1,0){4}} 
\put(0,0){\line(0,1){2}}  \put(2,0){\line(0,1){2}}
\put(4,0){\line(0,1){2}}
\multiput(5,1)(1,0){3}{\circle*{0.5}}
\put(8,0){\line(1,0){4}} \put(8,2){\line(1,0){4}} 
\put(8,0){\line(0,1){2}}  \put(10,0){\line(0,1){2}}
\put(12,0){\line(0,1){2}}
\multiput(1,3)(1,0){3}{\circle*{0.5}}
\multiput(5,3)(1,0){3}{\circle*{0.5}}
\multiput(9,3)(1,0){3}{\circle*{0.5}}
\put(0,4){\line(1,0){4}} \put(0,6){\line(1,0){4}} 
\put(0,8){\line(1,0){4}}
\put(0,4){\line(0,1){4}}  \put(2,4){\line(0,1){4}}
\put(4,4){\line(0,1){4}}
\put(0,6.5){\line(1,0){4}}
\put(0,7){\line(1,0){4}}
\put(0,7.5){\line(1,0){4}}
\multiput(5,7)(1,0){3}{\circle*{0.5}}
\multiput(5,5)(1,0){3}{\circle*{0.5}}
\put(8,4){\line(1,0){4}} \put(8,6){\line(1,0){4}}
\put(8,8){\line(1,0){4}}
\put(8,6.5){\line(1,0){4}}
\put(8,7){\line(1,0){4}}
\put(8,7.5){\line(1,0){4}}
\put(8,4){\line(0,1){4}}  \put(10,4){\line(0,1){4}}
\put(12,4){\line(0,1){4}}
\put(0.5,6){\line(0,1){2}}
\put(1,6){\line(0,1){2}}
\put(1.5,6){\line(0,1){2}}
\put(2.5,6){\line(0,1){2}}
\put(3,6){\line(0,1){2}}
\put(3.5,6){\line(0,1){2}}
\put(8.5,6){\line(0,1){2}}
\put(9,6){\line(0,1){2}}
\put(9.5,6){\line(0,1){2}}
\put(10.5,6){\line(0,1){2}}
\put(11,6){\line(0,1){2}}
\put(11.5,6){\line(0,1){2}}
\end{picture} 
 }_{N} \right) \le N h \left(  {\scriptstyle{m+1}} \Big\{ \; \begin{picture}(7,6) 
\put(0,-1){\line(1,0){6}}  \put(0,1){\line(1,0){6}}
\put(0,3){\line(1,0){4}}  \put(0,5){\line(1,0){2}}
\put(0,3.5){\line(1,0){2}}  \put(0,4){\line(1,0){2}} \put(0,4.5){\line(1,0){2}}
\put(0,-1){\line(0,1){6}} \put(2,-1){\line(0,1){6}}
\put(0.5,3){\line(0,1){2}} \put(1,3){\line(0,1){2}} \put(1.5,3){\line(0,1){2}} 
\put(4,-1){\line(0,1){4}} \put(6,-1){\line(0,1){2}}
\multiput(0,-2)(1,0){8}{\circle*{0.5}}
\end{picture}
\right) \; .
\label{inequality2}
\ee
For the sake of clarity we restricted
the argumentation to the unidirectional case. It will be obvious
how the same reasoning can be applied to the diffusive case. 

\begin{figure}[ht]
\centerline{\psfig{figure=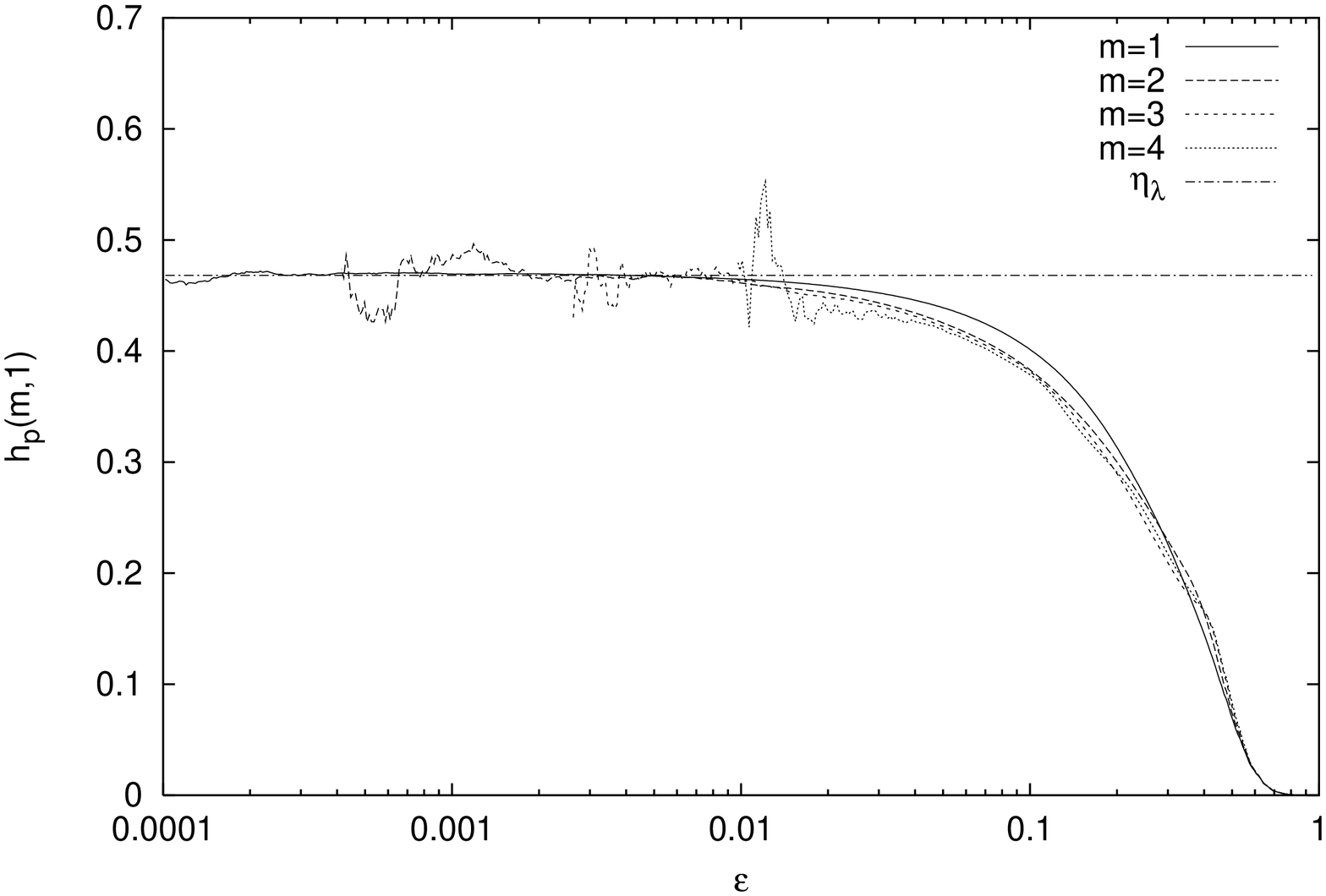,width=8cm,angle=270}
\psfig{figure=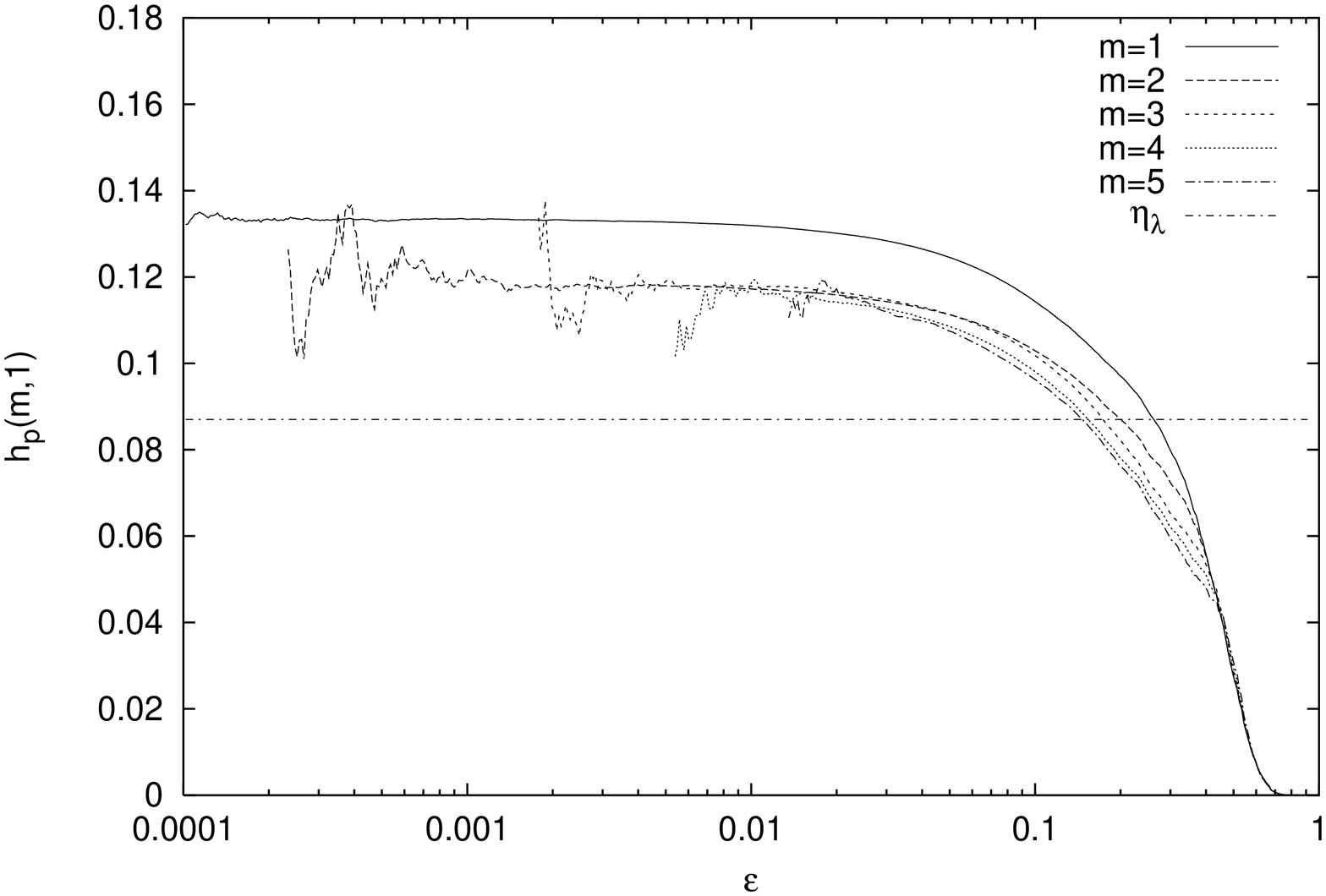,width=8cm,angle=270}}
\caption{Estimates of $h_p(m,1)$ for the unidirectional case,
linear coupling with $\sigma=0.2$ (left) and $\sigma=0.5$ (right). }
\label{uni-0.5}
\end{figure}

Let us introduce the abbreviation
\be
h_p(m,1):= h \left(  {\scriptstyle{m+1}} \Big\{ \; \begin{picture}(7,6) 
\put(0,-1){\line(1,0){6}}  \put(0,1){\line(1,0){6}}
\put(0,3){\line(1,0){4}}  \put(0,5){\line(1,0){2}}
\put(0,3.5){\line(1,0){2}}  \put(0,4){\line(1,0){2}} \put(0,4.5){\line(1,0){2}}
\put(0,-1){\line(0,1){6}} \put(2,-1){\line(0,1){6}}
\put(0.5,3){\line(0,1){2}} \put(1,3){\line(0,1){2}} \put(1.5,3){\line(0,1){2}} 
\put(4,-1){\line(0,1){4}} \put(6,-1){\line(0,1){2}}
\multiput(0,-2)(1,0){8}{\circle*{0.5}}
\end{picture}
\right) \; .
\label{pyramide}
\ee
The index $p$ means ''pyramid'' and denotes the form of the
spatio-temporal embedding. The first argument $m$ gives the number of
time steps used for prediction and the second argument denotes the number of
lattice sites predicted. That means e.g.
\[
h_p(2,3) = h \left( \begin{picture}(11,5)
\put(0,-2){\line(1,0){10}}  \put(0,0){\line(1,0){10}}
\put(0,2){\line(1,0){8}}  \put(0,4){\line(1,0){6}}
\put(0,2.5){\line(1,0){6}}  \put(0,3){\line(1,0){6}} \put(0,3.5){\line(1,0){6}}
\put(0,-2){\line(0,1){6}} \put(2,-2){\line(0,1){6}}
\put(4,-2){\line(0,1){6}} \put(6,-2){\line(0,1){6}}
\put(0.5,2){\line(0,1){2}} \put(1,2){\line(0,1){2}}
\put(1.5,2){\line(0,1){2}}
\put(2.5,2){\line(0,1){2}} \put(3,2){\line(0,1){2}}
\put(3.5,2){\line(0,1){2}}
\put(4.5,2){\line(0,1){2}} \put(5,2){\line(0,1){2}}
\put(5.5,2){\line(0,1){2}}
\put(8,-2){\line(0,1){4}} \put(10,-2){\line(0,1){2}}
\end{picture}
\right)\;.
\]
\end{multicols}
\begin{figure}[t] 
\centerline{     
\psfig{figure=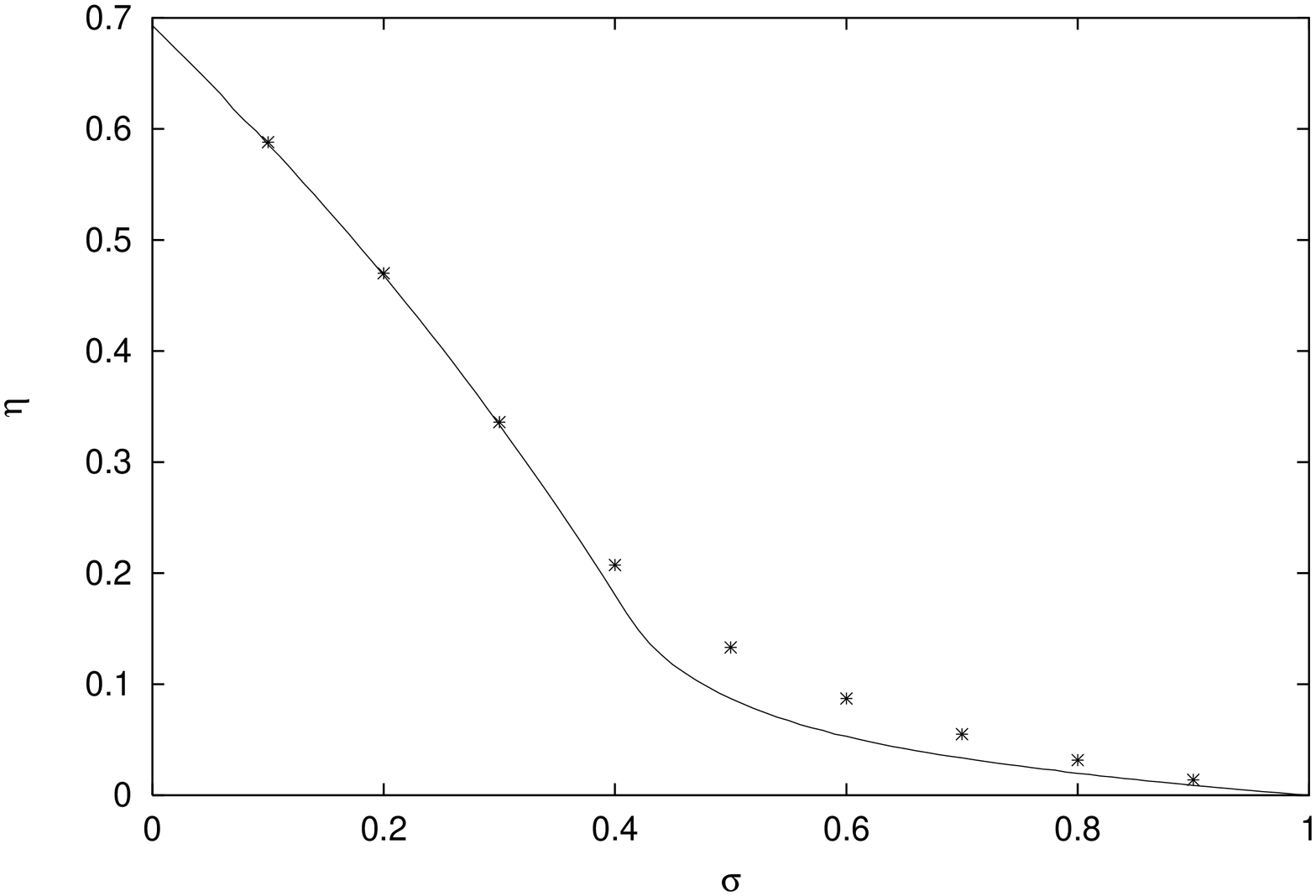,width=8cm,angle=270}
\psfig{figure=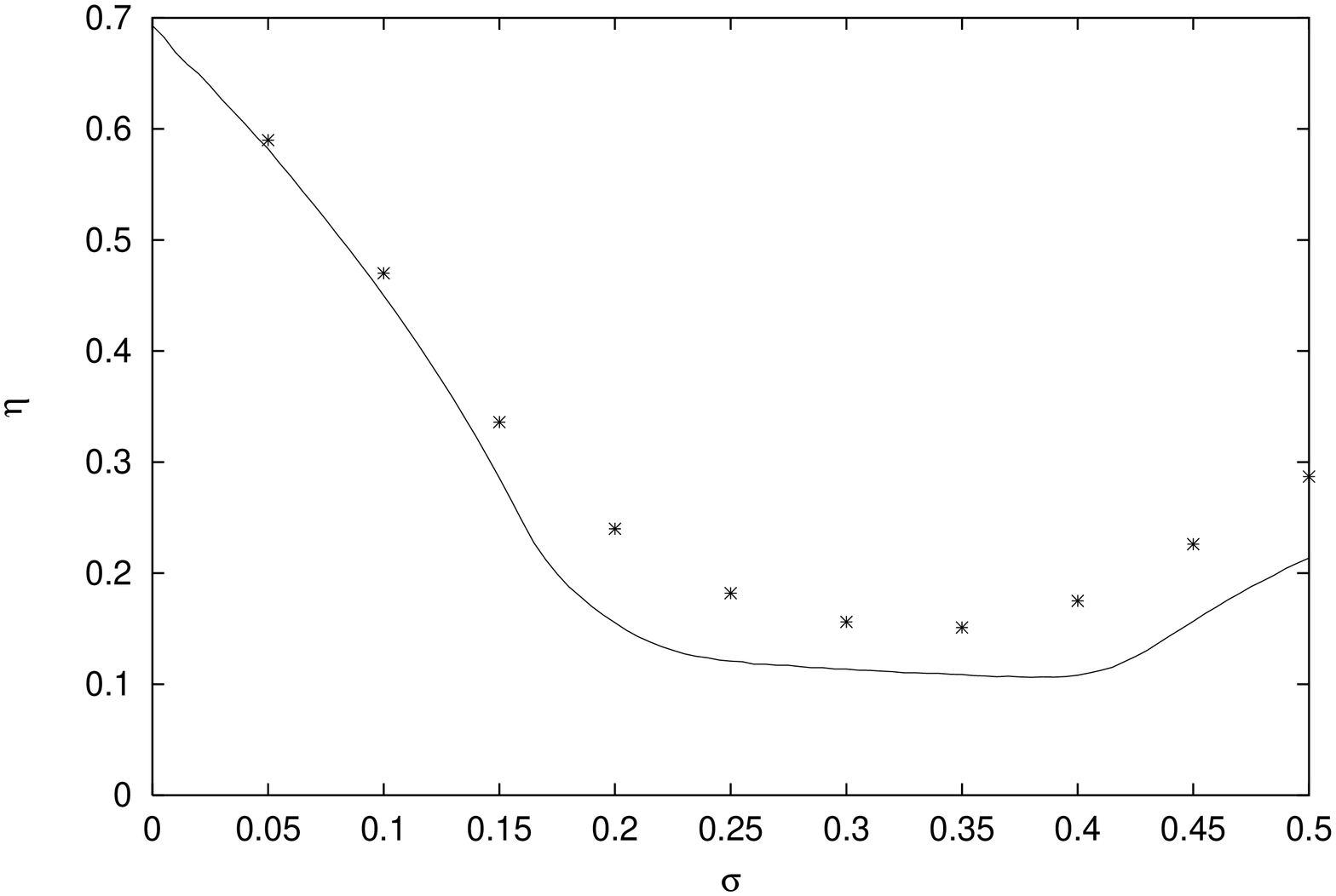,width=8cm,angle=270}
}
\caption{Estimates of $h_p(1,1)$ (stars) for the unidirectional linear coupling
(left) and the diffusive nonlinear one (right) compared to the entropy
density computed via the Pesin identity, $\eta_{\lambda}$.}
\label{ent-den}
\end{figure}
\begin{multicols}{2}
Now we can formulate an upper bound for the entropy density:
\be
\eta \le \lim_{\epsilon \to 0} \frac{1}{n} h_p(m,n) \;.
\label{bound}
\ee
Here we used again the same type of inequality like (\ref{inequality})
which leads to  \bes
h_p(m,n) \le h_p(m',n)  \qquad &\mbox{if}& \qquad m > m' \;.
\label{h-m-inequality} \\
\frac{1}{n} h_p(m,n) \le \frac{1}{n'} h_p(m,n')  \qquad &\mbox{if}& \qquad n > n' \;.
\label{h-n-inequality}
\ees
Similar to the usual conditional entropies in low dimensional systems
the entropies $h_p(m,n)$ become constant on sufficiently small length
scales (see e.g. Fig.~\ref{uni-0.5}). Thus the entropies $h_p(m,n)$
provide an upper bound for the entropy density already on finite
length scales $\epsilon$. If one is interested in the quality of the
bound one has to consider the inequalities~(\ref{h-m-inequality}),
which reflect the effects of correlations in time,
and~(\ref{h-n-inequality}), which reflect the spatial
correlations. Generally, the tighter the bound
(\ref{inequality}) should be, the less local the measurement has to
be. 

While the results in the former section were derived for the usual
Shannon entropy, for numerical investigations the entropies based on
the correlation sum are much more convenient. They provide better
statistics and require less computational effort.  The correlation sum
is defined as 
\be 
C(\epsilon)=\frac{1}{N (N-1)} \sum_{i \neq j}
\Theta(\epsilon-|\vec{s}_i-\vec{s}_j|) \; .  
\ee 
The quantity
$H(\vec{s},\epsilon)=-\ln C(\epsilon)$ can be regarded as a
generalized entropy, the so called correlation entropy
\cite{Grassberger83b,Grassberger84a,Takens98}.  One disadvantage of
using the correlation entropy is that inequality (\ref{inequality})
does no longer hold rigorously. Although experience shows that 
the deviations are usually small, to our knowledge there do not exist
theoretical arguments supporting this. A second disadvantage is that the
correlation entropy is a lower bound of the Shannon entropy. Thus,
strictly speaking we cannot expect to estimate an upper bound of
the entropy density by using the correlation integral. Nevertheless,
we can interpret the results provided by the correlation sum as
approximate estimates of the entropy density as it was done
with the usual correlation entropies as approximate estimates for the
KS-entropies of low dimensional systems
(e.g. \cite{Grassberger83b}).

\begin{figure}[ht]
\centerline{\psfig{figure=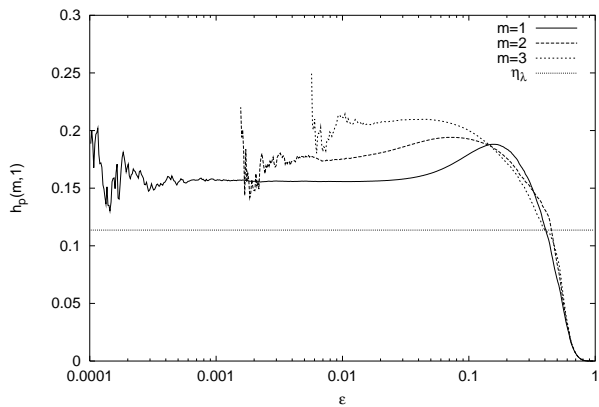,width=8cm,angle=270}}
\caption{$h_p(m,1)$ $m=1,2,3$ for the diffusive case,
$\sigma=0.3$  estimated by using the correlation entropies.}
\label{compare-bi-0.3}
\end{figure}

Fig.~\ref{ent-den} shows the
estimates of $h_p(1,1)$ as a function of the coupling $\sigma$
estimated by using the correlation sum (stars).
They are compared to the results for the entropy density calculated by
the Pesin identity (solid line) via the Lyapunov exponents
$\eta_{\lambda}=1/N \sum_i \lambda_i$ with $\lambda_i >0$.
The scaling
with respect to $\epsilon$ can be seen in Fig.~\ref{uni-0.5} for
the unidirectional coupling with $\sigma=0.2$ and $\sigma=0.5$. In the
example with $\sigma=0.2$ the estimates of $h_p(m,1)$ are almost independent
of $m$ and turn out to be a good approximation of the entropy density,
calculated by the Pesin-identity. This coincides with the
observation in Fig.~\ref{ent-den} that the value of $h_p(1,1)$ is very
close to the value calculated via the Lyapunov exponents for $\sigma
\le 0.3$ for the unidirectional and $\sigma \le 0.05$ in the diffusive
case. For larger coupling effects of correlations become visible. In
Fig.~\ref{uni-0.5} one sees that $h_p(2,1)$ is a remarkably better
estimate than $h_p(1,1)$. Further increasing $m$ gives no
better results. The remaining difference between $h_p(m,1)$ and the Pesin
value might be due to the spatial correlations in the system.
\end{multicols}
\begin{multicols}{2}
Moreover, in the strong coupling case the problem of the violation
of (\ref{inequality}) by the correlation entropies becomes
relevant. Fig.~\ref{compare-bi-0.3} shows $h_p(m,1)$ for $m=1,2,3$. As
one can see the inequality (\ref{h-m-inequality}) is strongly
violated. The reason for this behaviour is still an open question.

In summary, we presented a method to estimate the entropy density in
coupled map lattices with local couplings using only observables of
local subsystems which corresponds to the estimates of the KS-entropy
in low dimensional systems. As shown in
eqn.~(\ref{eta2})-(\ref{inequality2}) this would be an upper bound of
the KS--entropy, if we could use an algorithm based on a partition of
the phase space. Unfortunately, this is not possible due to the
statistical requirements. The correlation method uses a covering,
instead, so that the inequalities shown are no longer valid,
rigorously. Nevertheless, the numerical investigation we presented for
two simple cases showed that this method provides a rather good
approximation of the entropy density.

We would like to thank Peter Grassberger for stimulating discussions
and for carefully reading the manuscript.

\end{multicols}





\begin{thebibliography}{10}
\bibitem{footnote}
$h(xy|z)=H(xyz)-H(z)$. Adding the terms
$H(yz)-H(yz)=0$, we get $h(xy|z)=H(xyz)-H(yz)+H(yz)-H(z)=h(x|yz)+h(y|z)$.
\end{thebibliography}
\end{document}